\documentclass[preprint]{aastex}
\usepackage{epsf}

\shorttitle{Analysis Sunspot Torsional Oscillation}
\title{Analysis of \textit{SOHO}/MDI and \textit{TRACE} Observations of Sunspot Torsional Oscillation in AR10421}
\author{O.S.~Gopasyuk$^{1}$ and
        A.G.~Kosovichev$^{1,2}$
       }
\affil{$^{1}$ Crimean Astrophysical Observatory, Nauchny, Crimea 98409, Ukraine\\
           $^{2}$ Stanford University, Stanford, CA 94305, USA}
\altaffiltext{1}{email:olg@crao.crimea.ua}
\begin{document}

\begin{abstract}
Rotation of the leading sunspot of active region NOAA 10421 was investigated using magnetograms and Dopplergrams from the MDI instrument of the \textit{Solar and Heliospheric Observatory} (\textit{SOHO}), and white-light images from the \textit{Transition Region and Coronal Explorer} (\textit{TRACE}). The vertical, radial, and azimuthal axisymmetrical components of both magnetic and velocity field vectors were reconstructed for the sunspot umbra and penumbra. All three components of both vectors in the umbra and penumbra show torsional oscillations with the same rotational period of about 3.8 days. The \textit{TRACE} white-light data also show that the sunspot umbra and penumbra are torsionally rotating with the same period. Possible mechanisms of sunspot torsional motions are discussed.
\end{abstract}


\section{INTRODUCTION}

Sunspot rotation is one of the significant and long-standing problems of solar magnetism. It is related to the origin and evolution of sunspots and active regions. Rotation of sunspots around their centers has been observed by many authors \citep{Abetti,Maltby,Gop65,Gop81,Gop82,Brown01,Brown03,Zhao2003,Night,GopGop05}. Formation and strengthening of the vortical structure of the transverse magnetic field of sunspots  correspond well to the sunspot rotation calculated from photoheliograms \citep{Gop65}. Sunspot rotation is often accompanied by flares \citep{Gop65,GopLaz86,Brown03,Zhang}.

The torsional motions of sunspots with a mean period of about six days were found from studies of photoheliograms, line-of-sight velocity in the photosphere, and $H_{\alpha}$  images of active regions \citep{Gop81,Gop82}. Further investigations showed that such motions are rather common and led to their interpretation in terms of sunspot torsional oscillations, which may represent an intrinsic feature of the sunspot state \citep{GopLyam87}.

The determination of sunspot rotation from measurements of line-of-sight velocity over long periods of time is a laborious task \citep{Kin52,Lamb,Gop77}. Therefore, this investigation has only been carried out in a few cases. A method allowing determination of all three components of the field vector using its line-of-sight component \citep{GopGop98} significantly simplified the problem and expanded possibilities for studying the process of sunspot rotation. This method is based on the assumption that the primary sunspot structure is axisymmetric. The magnetic field in the sunspot penumbra is mainly horizontal, whereas in the umbra it is basically vertical. Therefore, variations in the penumbra and umbra of sunspots are investigated separately.

In this paper, we investigate torsional oscillations of the leading sunspot of AR 10421 using observations in the photosphere of both velocity and longitudinal magnetic fields and white-light images. The torsional oscillations of these sunspots were first noticed in the \textit{Transition Region and Coronal Explorer} (\textit{TRACE}) images by \citet{Night}. The simultaneous
observations of the sunspot in the high-resolution field of view of the \textit{Solar and Heliospheric Observatory} (\textit{SOHO})/MDI instrument \citep{Scherrer1995} and \textit{TRACE} allowed us to compare the oscillation characteristics determined from the magnetograms and Doppler-shift data with the oscillations of the sunspot structure in white light.

\section{OBSERVATIONS AND DATA REDUCTION}
The \textit{SOHO}/MDI magnetograms and Dopplerograms in the Ni~{\sc i} $\lambda$6768 absorption line and continuum images of the leading sunspot of active region NOAA 10421 with a spatial resolution of 0.6 arcsec and a temporal cadence of 1 minute were used. The data cover the time interval between 22:00 UT, 2003 August 1, and 21:00 UT, 2003 August 3. The sunspot was in the southern hemisphere. The latitude and longitude of the sunspot are presented in Table~\ref{tab1}. For a more detailed study of sunspot rotation, the corresponding white-light images from \textit{TRACE} were also used.

\begin{figure}[ht]
\centerline{\hspace*{0.01\textwidth} \includegraphics[width=0.33\textwidth,clip]{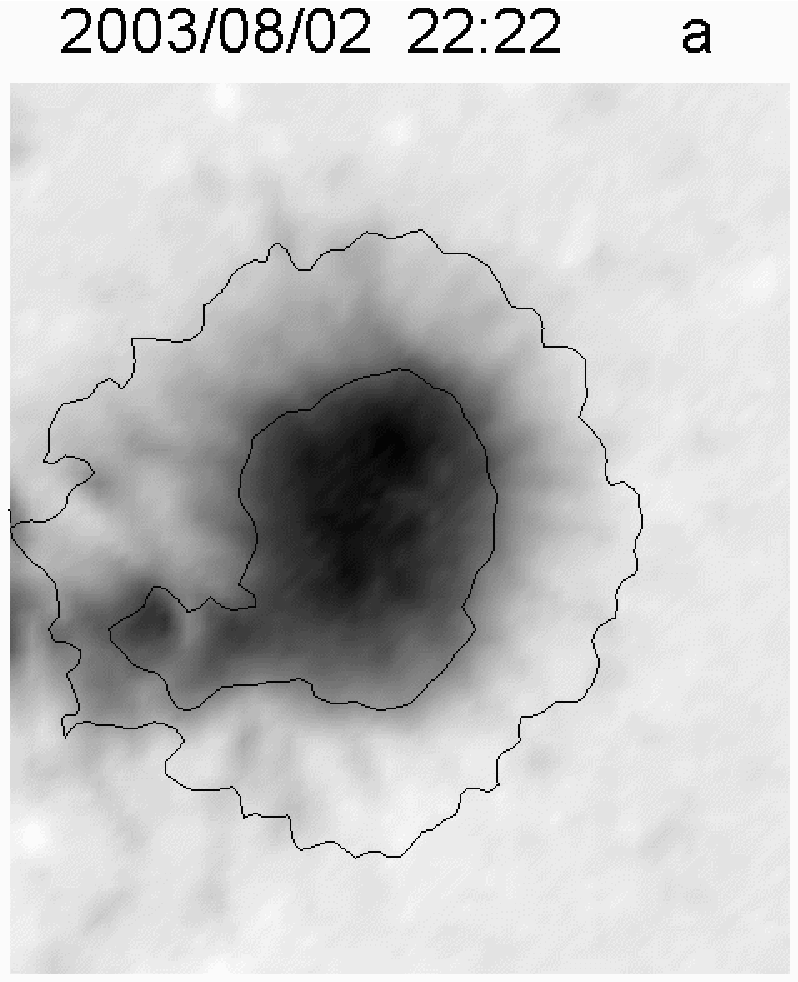}
\hspace*{-0.02\textwidth} \includegraphics[width=0.33\textwidth,clip]{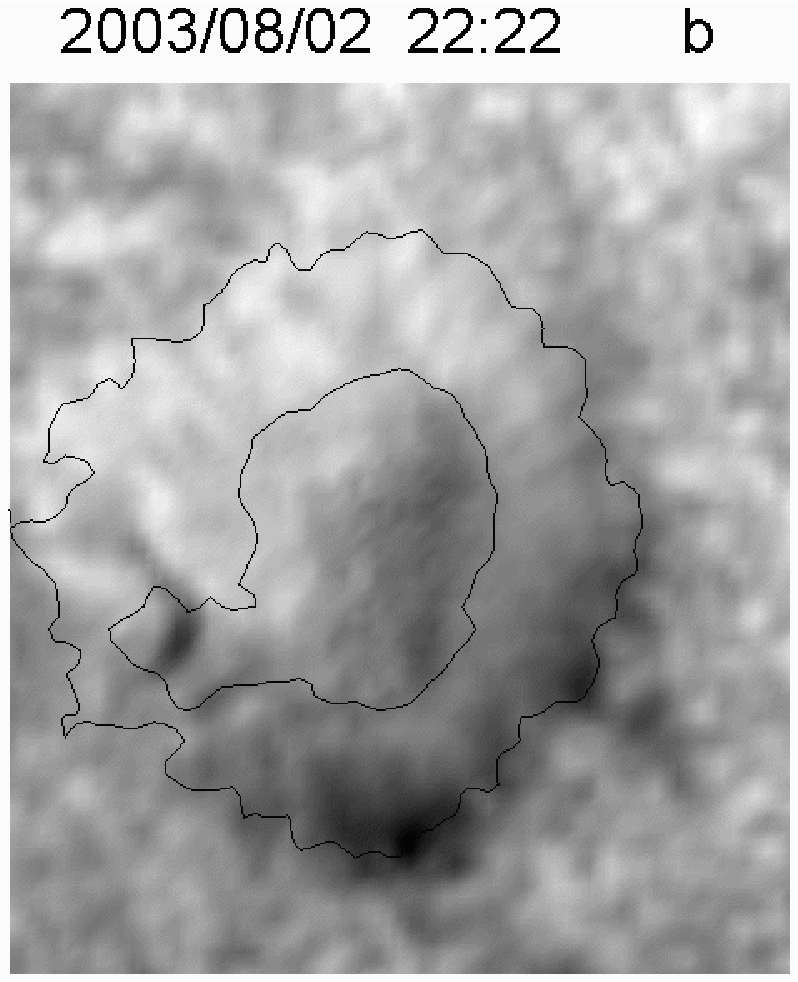}
\hspace*{-0.02\textwidth} \includegraphics[width=0.33\textwidth,clip]{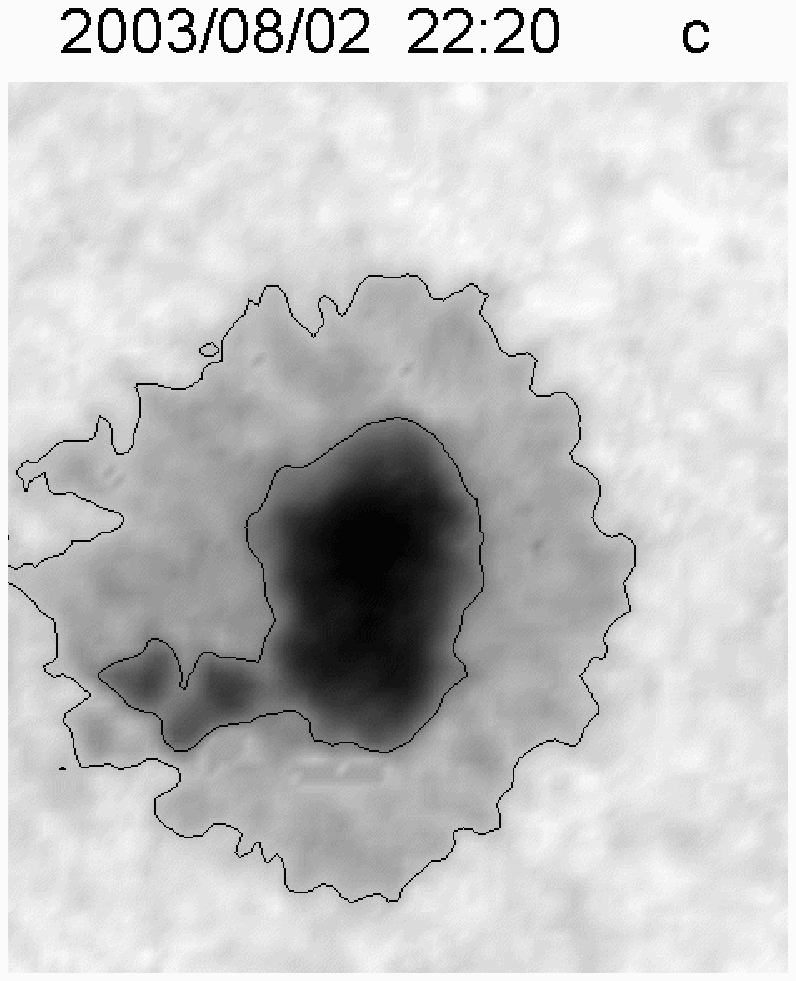}}
\caption{
Example of sunspot data of 2003 August 2, showing magnetogram (a), Dopplergrams from \textit{SOHO}/MDI (b), and photospheric white-light images from \textit{TRACE} (c). Solid lines outline the boundaries of the umbra ($0.7 \leq I$) and the penumbra ($0.7 < I \leq 0.9$).
}
\label{fig1}
\end{figure}

\begin{table}[h]
\caption{Observational Data and Properties of Torsional Oscillations of the Leading Sunspot of NOAA 10421 }\label{tab1}
\begin{tabular}{lllllll} \hline
\multicolumn{2}{l}{Date}& \multicolumn{5}{c}{1.08.2003 -- 3.08.2003}\\
\multicolumn{2}{l}{Latitude}& \multicolumn{5}{c}{	S06}\\
\multicolumn{2}{l}{Longitude}& \multicolumn{5}{c}{	13E -- 17W}\\
\hline
\multicolumn{4}{c}{\textit{SOHO}/MDI}&\multicolumn{3}{c}{\textit{TRACE}}\\
\hline
\multicolumn{2}{c}{ }& \multicolumn{1}{c}{Umbra}& \multicolumn{1}{c}{Penumbra}&
\multicolumn{1}{c}{}&\multicolumn{1}{c}{Umbra}& \multicolumn{1}{c}{Penumbra}\\
\multicolumn{2}{l}{Radius}&$8^{\prime\prime}.2$ & $16^{\prime\prime}.1$ & $R$ & $7^{\prime\prime}.8$ & $14^{\prime\prime}.5$ \\
\multicolumn{2}{l}{$P$ (day)}&3.8&3.8&$P$ (day)&3.8&3.8\\
$V_f$&$A$ (m s$^{-1}$)&295.5&223.4&$V_p$ (m s$^{-1}$)&4.3&12.1\\
 &$\alpha$	(rad)&3.9&5.3&$\alpha$ (rad)&5.0&1.0\\
$V_r$&$A$ (m s$^{-1}$)&158.3&154.0& $A_0$ (deg)&14.2&21.6\\
 &$\alpha$ (rad)&3.6&2.1 & & &\\
$V_z$&$A$ (m s$^{-1}$)&63.9&103.5 & & &\\
 &$\alpha$ (rad)&3.9&3.7 & & &\\
$H_f$&$A$ (G)&227.5&176.9 & & &\\
 &$\alpha$ (rad)&3.6&2.1 & & &\\
$H_r$&$A$ (G)&288.1&60.9 & & &\\
 &$\alpha$ (rad)&5.1&1.1 & & &\\
$H_z$& $A$ (G)&56.7&34.9 & & &\\
 &$\alpha$ (rad)&1.6&0.4 & & &\\
\hline
\end{tabular}
\end{table}

The MDI continuum images were used to determine the average positions and boundaries of the sunspot umbra and penumbra. The continuum intensity, $I$, is expressed in units of photospheric brightness. Following \citet{BL}, the umbra and penumbra are defined as the regions where $I\leq 0.7$ and $0.7 <I\leq 0.9$, respectively.
Samples of the magnetograms, Dopplergrams, and white-light images are presented in Figure~\ref{fig1}.

\begin{figure}
\centerline{\includegraphics[width=12cm]{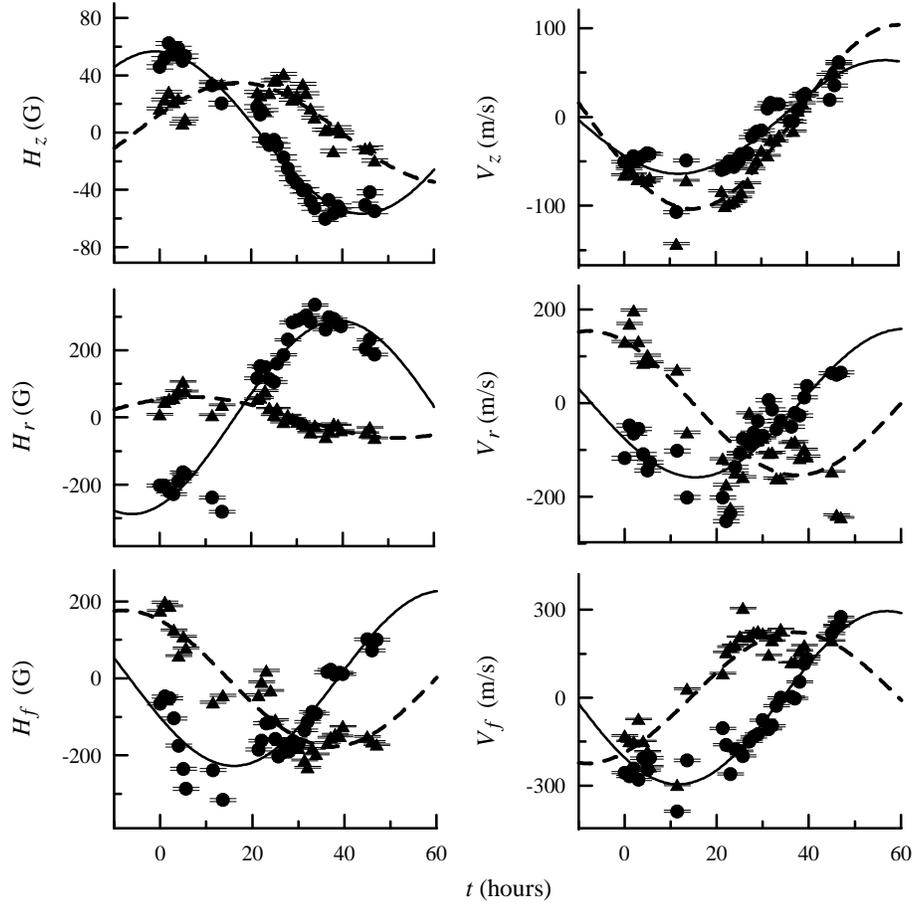}}
\caption{Dependences of the magnetic and velocity field components on time for the umbra (circles) and the penumbra (triangles) of the NOAA 10421 leading sunspot. The data are fitted by sinusoidal curves.} \label{fig2}
\end{figure}

Our study was based on the method of restoration of all three components of the magnetic and velocity field vectors using their line-of-sight components under assumption of axial symmetry \citep{GopGop98,GopGop05}.
Separately for the umbra and the penumbra, we calculated three components of magnetic $\vec{H}$ and velocity $\vec{V}$ field vectors for each magnetogram and Dopplergram: vertical $H_z$ ($V_z$), radial $H_r$ ($V_r$) (along the sunspot radius), and azimuthal $H_f$ ($V_f$). Variations of the averaged values of the components are plotted as a function of time $t$ in Figure~\ref{fig2}. Time $t= 0$ corresponds to 22:00 UT, 2003 August 1. Each of the magnetic and velocity components was fitted with a sinusoidal function:
\begin{equation}
y = A \cdot\sin\left(\frac{2\pi t}{P} + \alpha
\right), \label{eq1}
\end{equation}
where $A$, $P$, and $\alpha$  are the amplitude, period, and initial phase respectively. Their values are shown in Table~\ref{tab1}. The fits shown in Figure~\ref{fig1} provide evidence that sunspot rotation is characterized by torsional oscillations of the sunspot magnetic structure. The period of oscillations for all components of the magnetic and velocity fields in the umbra and penumbra  is the same, $\approx 3.8$ days. However, the amplitude and phases are different. In particular, the phase of the velocity azimuthal component in the umbra and penumbra is ahead of the phase of the magnetic field azimuthal component.

\begin{figure}
\centerline{\includegraphics[width=7cm]{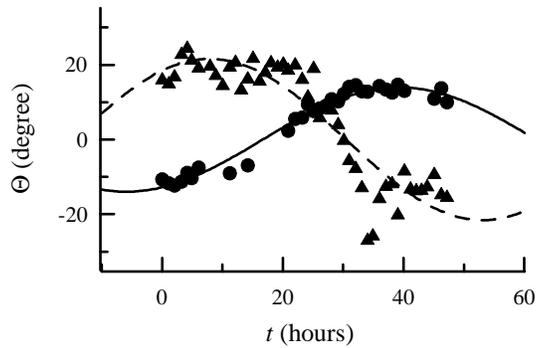}}
\caption{Temporal dependence of the angles of rotation of the umbra (circles) and the penumbra (triangles) of NOAA 10421 sunspot from \textit{TRACE} white-light images. The data approximated sine curves are also shown.} \label{fig3}
\end{figure}
The period of the sunspot's rotation about its center was also determined using white-light images of the sunspot. We considered the center of ``mass" of these images to be the center of the sunspot. In every image, a ``sunspot axis", the line passing through the sunspot center and the center of a characteristic detail of the umbra (or penumbra), which remained during the whole observational period, was determined. The angle of sunspot rotation, $\Theta$,  is defined as the angle between the ``sunspot axis" and the solar equator line. The values of $\Theta$ averaged over an hour were calculated. Figure~\ref{fig3} shows the changes of the angle of rotation of the umbra and the penumbra with time. These temporal variations have been fitted by sinusoids (Equation~(\ref{eq1})). The period, amplitude, and phase of these fits are presented in Table~\ref{tab1}. The period of the umbra and penumbra oscillations from the white-light images is also 3.8 days.

Rotation of the leading sunspot of NOAA 10421  was previously investigated by \citet{Night} from \textit{TRACE} observations. \textit{TRACE} observed this active region in white light, ultraviolet, and extreme ultraviolet passbands. Analysis of the photospheric white-light images showed that the leading sunspot was torsionally rotating about the center of its umbra. Twisted coronal fans can be seen above the rotating sunspot in the EUV images on and after August 1.

\section{Discussion}

Currently, the nature of the twisting motions of sunspots is not understood.
If one assumes that the sunspot rotation derived from white-light observations represents a solid-body rotation of the whole spot, then the azimuthal velocity of the sunspot rotation is defined as
\begin{equation}
V_{p}= 2\pi \frac{A_0}{360} \frac{R}{P},
\label{eq2}
\end{equation}
where $P$ is the period of sunspot rotation, $R$ is an average radius of characteristic details of the umbra (or penumbra) from the sunspot center (Table~\ref{tab1}), and $A_0$ is the mean amplitude of the rotation angle. The estimates of $V_p$ (in m s$^{-1}$) are shown in Table~\ref{tab1}. Apparently, the oscillatory variations of the azimuthal velocity calculated from the Dopplergrams in the umbra exceed the  azimuthal speed measured from the white-light data by a factor of approximately 60. The same is true of those calculated for the penumbra. Thus, the sunspot rotation and torsional oscillatory motions are not of a solid-body or a single flux-tube type, but must involve a complex interplay between internal dynamics and the global sunspot structure.

\begin{figure}[h]
\centerline{\hspace*{0.01\textwidth} \includegraphics[width=0.4\textwidth,clip]{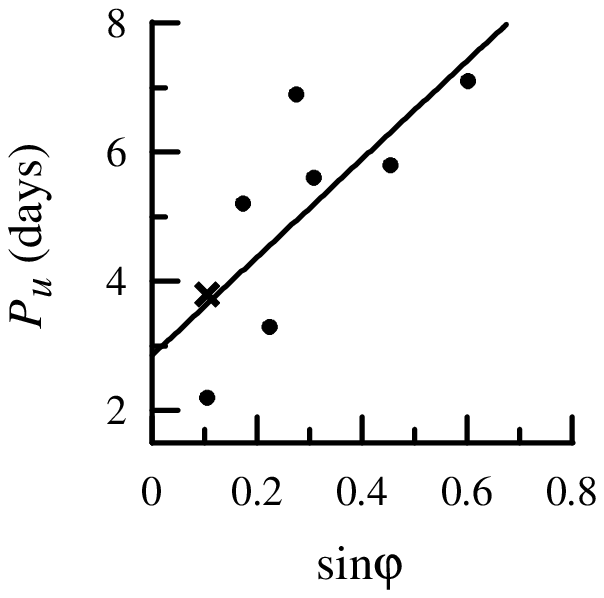}
\hspace*{0.05\textwidth}\includegraphics[width=0.4\textwidth,clip]{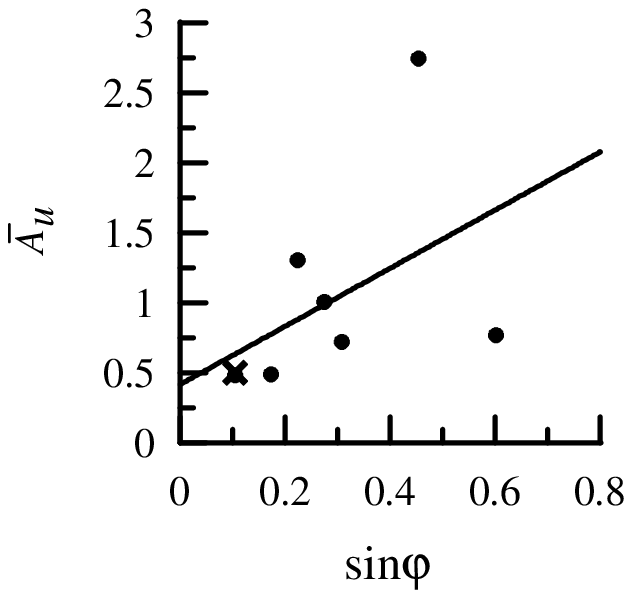}}
\caption{Period (left) and relative amplitude (right) of torsional oscillations of sunspot umbra as a function of sine latitude for previously studied sunspots  
(dots; Gopasyuk \& Gopasyuk 2006)
and the leading sunspot of NOAA 10421 (crosses).}\label{fig4}
\end{figure}

In general,
torsional oscillations in plasma with a magnetic field are generated by the forces of tension of the magnetic field lines trying to return the displaced plasma to its initial condition. As a result, a perturbation spreads along the magnetic field with the Alfv\'en wave velocity.
\citet{GopGop06} found that the characteristic period and amplitude of the torsional oscillations increase with sunspot latitude. Figure~\ref{fig4} shows this dependence for previously studied sunspots and the sunspot of NOAA 10421. This probably means that the torsional oscillations are forced oscillations. It has been proposed that their mechanism may be related to a joint action of surrounding supergranules. The azimuthal component of the Coriolis force acting on supergranular flows produces the azimuthal component of plasma velocity, which leads to twisting of magnetic field force lines. When the supergranular motions are occasionally synchronized, this may result in significant torque on the sunspot structure.
The estimates of \citet{GopGop06} give reason to suspect that this effect could be a reason for the torsional oscillations of sunspots on the solar surface.

The oscillatory motions of sunspots may also arise from the twisting of emerging flux tubes as demonstrated by numerical simulations \citep{Cheung2010}. However, one would expect that this mechanism plays a role only during flux emergence and not in mature sunspots with stable magnetic flux. Recent numerical simulations of \citet{Kitiashvili2010} show that the process of spontaneous formation of magnetic structures can be initiated by strong vortexes at the boundaries of granular cells. This may also result in twisting of the magnetic structures.

Local helioseismology observations have revealed strong diverging and converging flows beneath sunspots \citep{Kosovichev1996,Zhao2001,Zhao2010}, and also converging flows on a larger scale around active regions \citep{Haber2003,Zhao2004}. The subsurface flow patterns evolve with the evolution of sunspot regions. The time-distance helioseismology analysis showed predominant outflows during the emergence and decay phases, and converging inflows around developed sunspots \citep{Kosovichev2009}. The precise structure of these flows, their relationship to the Evershed and moat flows, and their role in the structure and dynamics of sunspots, are still being investigated \citep{Kosovichev2010}. However, it is intriguing that the helioseismology analysis found evidence of a vortex flow pattern changing with depth beneath a rotating sunspot \citep{Zhao2003}.

The current results encourage further investigation of the torsional oscillations of sunspots by using spectropolarimetric and helioseismology observations, and numerical simulations.

\section{CONCLUSION}
Using  observations of the line-of-sight magnetic field and velocity from \textit{SOHO}/MDI, and white-light images from \textit{TRACE}, we  studied the rotation of the leading sunspot of active region NOAA 10421. The vertical,  radial, and azimuthal  components of both magnetic and velocity field vectors were reconstructed for the umbra and penumbra of the sunspot, assuming axial symmetry. All three components reveal torsional oscillations. The study of the photospheric white-light images also shows that the leading sunspot was torsionally rotating about its center. In this particular case, the direction of the sunspot's twisting motion changed during the period of observations.

We determined the period, amplitudes, and phrases of these oscillations. The oscillation period estimated from the \textit{SOHO}/MDI and \textit{TRACE} data is $\approx 3.8$ days for both the umbra and the penumbra. The phase of the azimuthal component of velocity is ahead of the phase of the azimuthal magnetic field component. The oscillation parameters of the sunspot of NOAA 10421 are in line with the latitudinal dependence previously found for other sunspots (Figure~\ref{fig4}). This indicates that the torsional oscillations of sunspots are forced oscillations, which may be related to plasma motions beneath the sunspots or in the surrounding supergranulation cells, and to the action of the Coriolis force on these motions.

In all cases, there is a discrepancy, previously noticed by \citet{Gop81}, between the azimuthal velocity defined from the white-light images and the rotation velocity of sunspots calculated from line-of-sight velocity (Table~\ref{tab1}). The rotational velocity calculated from the white-light images is tens of times smaller than the azimuthal velocity defined from the line-of-sight velocity. At the same time, the twisting angle of the magnetic field lines located at the penumbra's external boundary corresponds to the angle of sunspot rotation calculated from the photoheliograms. These effects are likely related to the fine structure and dynamics of the sunspot, and to the irregular nature of the magnetic fields and flows in sunspots. Undoubtedly, these problems require further investigation.

The authors are grateful to the \textit{SOHO}/MDI and \textit{TRACE} teams for their open data policy. \textit{SOHO} is a project of international cooperation between ESA and NASA.

\end{document}